\documentclass[a4paper,10pt]{article}
\usepackage{geometry}
\usepackage{setspace}
\usepackage[utf8]{inputenc}
\usepackage{amsmath}
\usepackage[english]{babel}
\usepackage{indentfirst}
\usepackage{caption}
\usepackage{subcaption}
\usepackage{float}
\usepackage{graphicx}
\usepackage{authblk}  
\usepackage{xcolor}

\title{Light-matter interaction inside an optical cavity: A perspective}
\author{Anneswa Paul\footnote{anneswapaul@iisc.ac.in} and Upendra Harbola\footnote{uharbola@iisc.ac.in}\\
\it Department of Inorganic and Physical Chemistry\\ \it Indian Institute of Science, Bengaluru}
\date{}

\begin{document}

\maketitle

\begin{abstract}

Light-matter interaction inside an optical cavity and formation of polaritonic states have gained interest in the past decades as it has 
direct applications in many research fields. Different regimes of light-matter coupling have been studied using different approximations, one 
of which is rotating wave approximation(RWA), which is said to be valid for weak coupling and 
near resonant regimes. In this study, we have categorized the light-matter coupling into four 
regimes depending on the validity of the RWA as 
moderate, $\lambda/\omega_c\leq 0.1$, strong, $0.1 \leq \lambda/\omega_c\leq 0.5$, ultra-strong, $0.5 \leq\lambda/\omega_c\leq 1.0$ and 
deep-strong, $\lambda/\omega_c\geq 1.0$ 
coupling, where $\lambda$ is the coupling 
strength and $\omega_c$ is the cavity frequency. 
In experiments, vacuum Rabi-splitting has been 
observed which is a clear indication of 
formation of polaritonic states. It is a common misunderstanding that the cavity remains in 
vacuum state when the coupled system is in the ground state. Here we show that upon coupling, 
the cavity has non-zero excitation even in the ground state, which is not captured by RWA. 
In fact, RWA breaks down completely to predict the ground state properties as it fails to capture the interaction between the matter and the cavity field.

\end{abstract}

\section*{Introduction}
Interaction of light with matter is the fundamental process in nature. In fact, as believed \cite{NatureChem9,RebeccaPCCP2016},  the life itself is a consequence of the light-induced chemical reactions. 
Given its importance, it's natural to think of controlling the light-matter interaction to our advantage, for example, to steer a chemical reaction towards a 
desired product and to enhance or suppress the rate of a reaction. In order to achieve this, light pulses of different shapes  have been invented \cite{weiner2011ultrafast,jiang2018electrons,dey2020controlling}. In recent years, however, 
it has been recognized that the light trapped inside an optical cavity can also be used to manipulate chemical reactions and alter matter properties 
\cite{hutchison2012modifying,thomas2016ground}. In fact, vacuum fluctuations inside the cavity 
have been shown to couple to molecular resonances to produce light-matter hybrid states \cite{genet2021inducing,maissen2014ultrastrong,forn2017ultrastrong}. This is particularly interesting as these new hybrid states, called polaritonic states, 
carry properties of  both the light (electromagnetic field) and the matter \cite{BalaNatureMat2023}. 
The generation of polaritonic states is a consequence of strong interaction which, 
in turn, is a result of the modified field density-of-states inside the cavity. Research in the cavity-induced strong light-matter coupling and its effects on chemical reactions 
has become popular over the last decade. 

Different regimes of the coupling, that is, moderate, strong, ultra-strong, and deep-strong, have been discussed in the literature
 \cite{casanova2010deep, larson2021jaynes}.
However, there is no discussion on the qualitative distinction between these different regimes. It appears to the authors that there is some misunderstanding 
regarding the hybrid nature of the  polaritonic states, even though some rigorous theoretical works are available in the literature \cite{mandal2023theoretical}. The aim of this work is to alleviate some of these confusions and also to comment on the nature and possible quantification of cavity field-matter coupling strengths. We will keep the discussion to a pedagogical level and refer to the literature for technical details. 

\section*{Light-matter interaction}

Assuming that the matter (molecule) can be approximated by an electric dipole, the interaction of the electromagnetic field and matter can be expressed by the following Hamiltonian, $H=H_M+H_F+H_I$, where $H_M, H_F,$ denote isolated molecule and field, and $H_I$ is the interaction between the two \cite{Craig-ThiruBook}.
\begin{eqnarray}
H_M&=& \sum_i \frac{p_i^2}{2m} +V(r) \nonumber\\
H_F&=& \frac{\epsilon_0}{2}\int d^3r [D^2(r,t)+c^2B^2(r,t)] \nonumber\\
H_I&=& \int d^3r \mu(r,t)\cdot D(r,t), 
\end{eqnarray}
where $p_i$ is the kinetic momentum of bound charges on the molecule and $V(r)$ is the total potential energy at point $r$, $\epsilon_0$ is the permittivity of the 
free space, $D(r,t) [B(r,t)]$ represent displacement electric [magnetic] field, respectively, and $\mu(r,t)$ is the molecular dipole which couples to the 
displacement field. Each of the above Hamiltonian can be quantized using standard method \cite{Craig-ThiruBook,Scully1997}.  

Considering only two quantized states, ground $|g\rangle$ 
and excited $|e\rangle$, of the molecule which is coupled to the single mode of the radiation field inside the cavity. The total Hamiltonian can be re-written in the quantized form as,
\begin{eqnarray}
\label{full-hamil}
H= \hbar\omega_1 |g\rangle\langle g| +\hbar\omega_2 |e\rangle\langle e| + \hbar \omega_c (a^\dag a+1/2) + \hbar\lambda (|e\rangle\langle g|+|g\rangle\langle e|)(a+a^\dag) 
\end{eqnarray} 
where operator $a (a^\dag)$ destroys (creates) the cavity field mode of frequency $\omega_c$, and the coupling strength $\lambda=|\eta E_0\mu_{eg}|/\hbar$ depends on the 
field amplitude $E_0$, refractive index, $\eta$,  and the molecular transition dipole matrix element between the ground and the excited state, $\mu_{eg}$. 
Here we have assumed that both the field polarization and the dipole are aligned in the same direction. This is the well known Jaynes-Cummings (JC) Hamiltonian \cite{jaynes1963comparison}.
 
The JC model is the most studied model in quantum optics as it is applicable for many interesting two-level systems, for example, transitions in spin and resonant 
atomic systems. A natural choice of basis which can be used to construct the eigenbasis of $H$ are the product states $|g,n\rangle=|g\rangle\otimes|n\rangle$ and $|e,m\rangle=|e\rangle\otimes|m\rangle$, where $|n\rangle$ and $|m\rangle$ are the number basis for the cavity-field state with $n$ and $m$ number of photons. The Hilbert space of $H$ is factorized into two isolated 
Hilbert spaces of fixed parity determined by the even or odd number of excitations in the combined molecule-field space \cite{casanova2010deep}.  That is, for example, if the molecule and the cavity-field both 
are in the ground states (zero excitation), $|g,0\rangle$, the system (cavity field and 
molecule) can change to 
$|e,1\rangle$, the state with two excitations. Thus energy of the system changes from $\hbar \omega_1$ to  $\hbar(\omega_2+\omega_c)$. This extra energy is 
supplied by the coupling $\lambda$. Of course, neither $|g,0\rangle$ nor $|e,1\rangle$ is an eigenstate of $H$ and therefore the system evolves in time into a state given by the 
linear combination (wavepacket) of all even parity states. Clearly, the weight of various (even) parity states is determined by the coupling $\lambda$. 
If $\lambda$ is extremely small (we shall quantify it later), the system will mainly stay in $|g,0\rangle$ state with negligibly small amplitude in other states, so that for all 
practical purposes, system remains in $|g,0\rangle$ which can now be approximated as the ground state of $H$. This situation is often made exact by invoking the so-called 
rotating-wave-approximation (RWA)\cite{Scully1997, larson2021jaynes}. 

The RWA effectively modifies the coupling term in the Hamiltonian. 
Consider the time-evolution operator: $U(t,0) =e^{-iHt/\hbar}$, which, using interaction picture with respect to the non-interacting Hamiltonian, 
$H_0=H_M+H_F$, can be recast as: $U(t,0)=e^{-iH_0t} 
\exp\left\{\frac{-i\lambda}{\hbar}\int_0^t d\tau (|e\rangle\langle g|e^{i\omega_{21}\tau}+|g\rangle\langle e|e^{-i\omega_{21}\tau})(a e^{-i\omega_c\tau}+a^\dag e^{i\omega_c \tau})\right\}$. Note that the integrand inside the second exponential contains four terms, two terms oscillate with frequency $(\omega_{21}-\omega_c)$ and the other two with 
frequency $(\omega_{21}+\omega_c)$, where $\omega_c>0$ and $\omega_{21}=\omega_2-\omega_1>0$. Near the resonance $\Delta=\omega_{21}- \omega_c\sim 0$, the integral involving 
frequency terms $\Delta$ dominate over those with frequency $(\omega_{21}+\omega_c)\sim 2\omega_c$. Thus near the resonance, one can ignore the 
high frequency terms and keep only the slow varying terms. This leads to the Hamiltonian within RWA where the coupling term contains only 
$\lambda(|e\rangle\langle g| a+|g\rangle\langle e|a^\dag)$. 
\begin{eqnarray}
\label{RWA-hamil}
H_{RWA}= \hbar\omega_1 |g\rangle\langle g| +\hbar\omega_2 |e\rangle\langle e| + \hbar \omega_c (a^\dagger a+1/2) 
+ \hbar\lambda(|e\rangle\langle g| a+|g\rangle\langle e|a^\dagger).
\end{eqnarray}
Note that with this approximation, the parity chain is now broken, that is, a state with $n$-excitations remains in the subspace of parity where the total number of excitations 
\begin{figure}[h]
    \centering
    \includegraphics[width=0.5\linewidth]{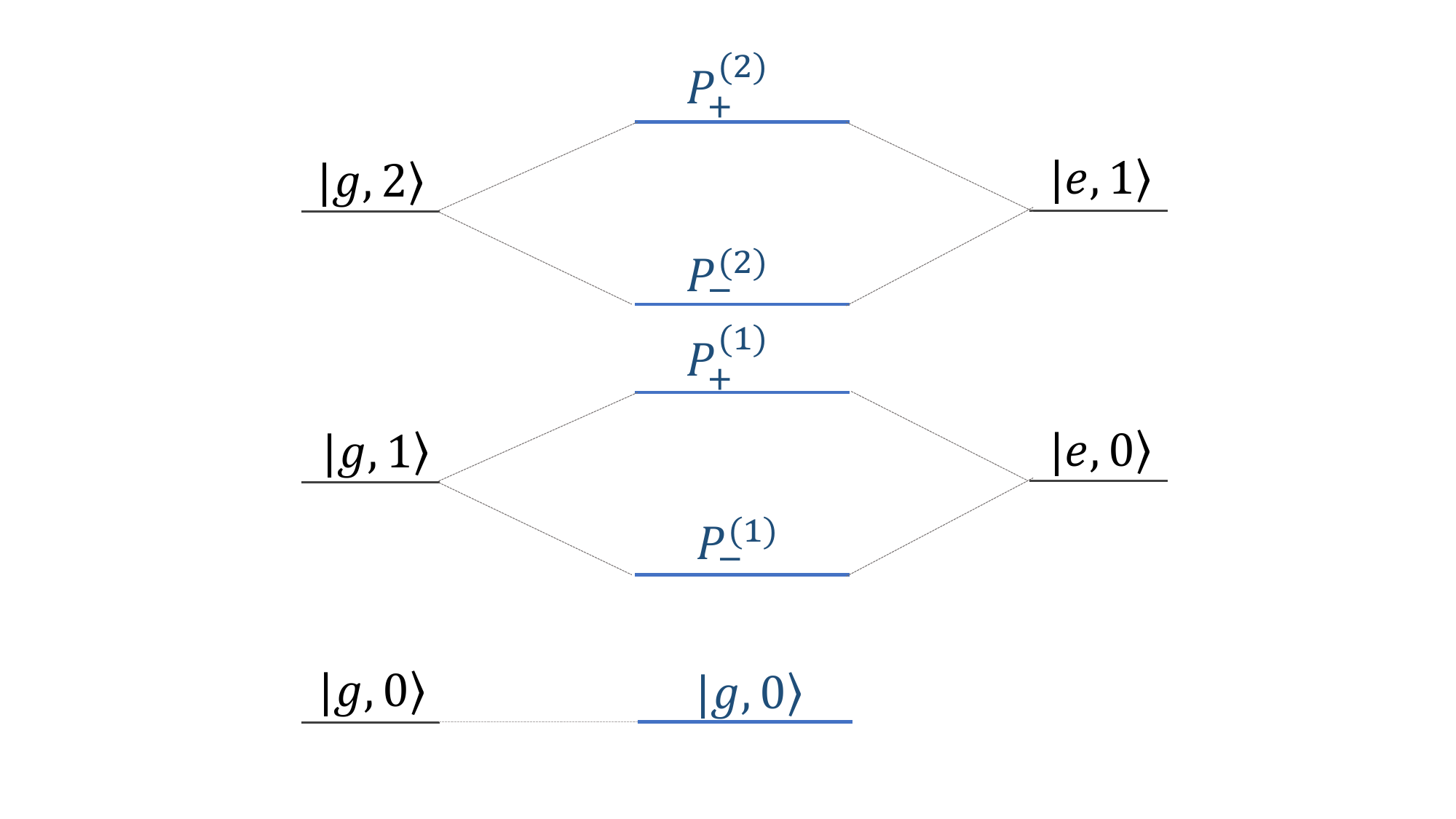}
    \caption{Schematic representation of the eigenspectrum of of $H_{RWA}$ at resonance, $\Delta=0$ upto the lowest three parity blocks, $|g,0\rangle$ and $P_{\pm}^{(n)}, n=1,2$, with zero, one, and two excitations in the system.}
    \label{fig:1}
\end{figure}
do not change. For example, the state with a single excitation, $|g,1\rangle$ can only couple to $|e,0\rangle$ and not to $|e,2\rangle$, which was possible under the full Hamiltonian, Eq.(\ref{full-hamil}). Thus the ground state (even parity, zero-excitation state) $|g,0\rangle$ is decoupled from all states and, therefore, is an eigenstate of $H_{RWA}$. 
Hence, if the matter and the cavity both are in their ground states, the two will not couple and no dynamics will take place.   This is the most fundamental
difference between the two Hamiltonians.

\begin{figure}[h!]
    \centering
    \includegraphics[width=0.5\linewidth]{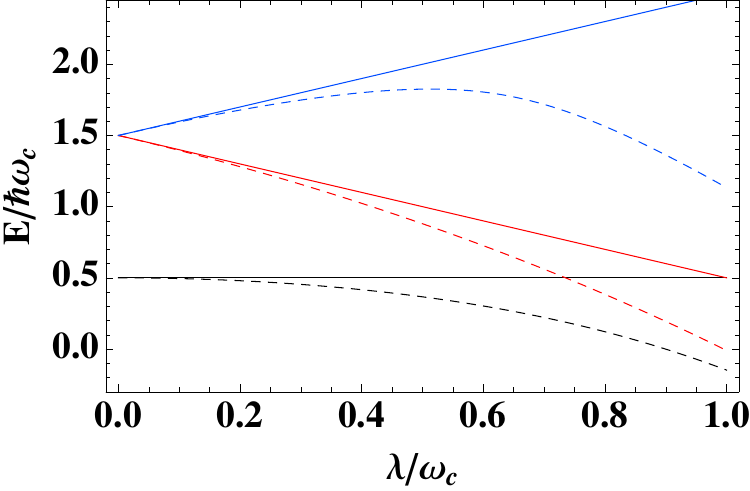}
    \caption{The eigen spectrum of $H_{RWA}$ (solid lines) is plotted as a function of the coupling strength, $\lambda$. The ground state, $|g,0 \rangle$ (black), and only the first-pair of polaritonic states, $P_-^{(1)}$ (red) and $P_+^{(1)}$(blue), are shown at the resonance condition. The lowest three eigenstates of the full Hamiltonian, $H$, are shown by black, red and blue dashed lines. Here $\omega_1=0$, $\omega_2=1$. }
    \label{fig:2}
\end{figure}

The RW Hamiltonian, Eq.(\ref{RWA-hamil}), has been used widely because it is an analytically solvable model. The eigenspectrum of $H_{RWA}$ is given by $\varepsilon_{\pm}^{(n)}=\frac{1}{2} \Big[\hbar(\omega_1+\omega_2+(2n-1)\omega_c \pm \hbar\sqrt{\Delta^2+4n\lambda^2})\Big]$, where $\varepsilon_{\pm}^{(n)}$ is the energy of polaritonic state $P_{\pm}^{(n)}$ in the $n$th excitation block and the off-resonant parameter, $\Delta\ll (\omega_1+\omega_2+\omega_c)$. Thus, it is clear that within the RWA, the ground state $|g,0\rangle$ does not couple. 
However, one must keep in mind that it is an approximate model and has two major limitations: it is valid near the resonance and for the weak coupling as discussed above. 
Schematic of the energy spectrum for $H_{RWA}$ is shown in Fig.(\ref{fig:1}). States $|g,0\rangle$, 
$P_{\pm}^{(n)}$ denote the eigenstates of $H_{RWA}$ with energies given by $\varepsilon_{\pm}^{(n)}$ defined above. The energy difference between each pair of the polaritonic states within the same excitation block increases monotonously with the coupling $\lambda$ and the excitation number $n$. This is shown in Fig.(\ref{fig:2}). The original resonance, therefore, gives 
rise to the two new resonances for transitions between the ground state $|g,0\rangle$ and the polaritonic states 
$P_\pm^{(1)}$ which appear as distinct peaks in the absorption spectrum of the cavity; a hallmark of the 
polaritonic state formation in the cavity. This is sometimes {\em erroneously} termed as signature of strong light-matter interaction in vacuum cavity. We reiterate that, within RWA, the molecule in its ground state does not ``see" the vacuum 
cavity, and, therefore, the two do not interact. In order to form the polaritonic states, cavity or the molecule must be excited.   

 Eigenspectrum of the full Hamiltonian $H$ is obtained by numerically diagonalizing the Hamiltonian. For this, we choose the natural product basis $|g,n\rangle$, $|e,m\rangle$ defined above. Ideally, $H$ is infinite-dimensional and cannot be diagonalized numerically. However, for the $\lambda$ values we are considering, a $30\times 30$ matrix formed by basis vectors containing upto $15$ excitations, works fine. A comparison between the eigenspectra of the two Hamiltonians, $H$ and $H_{RWA}$, is shown in Fig.(\ref{fig:2}). 

It is clear from Fig.(\ref{fig:2}) that at smaller values of the coupling, $\lambda/\omega_c \le 0.1$, the two eigenspectra agree well.
Although in literature, $\lambda/\omega_c \simeq 0.1$ is considered to be ultra strong coupling regime \cite{larson2021jaynes} where RWA fails, but here we show that in this region, RWA is a valid approximation for eigenvalues. Since the dynamics is also governed by the eigenvalues, we therefore expect that both the static and the dynamic quantities are approximated well by RWA within this regime. We can identify this coupling range as ``moderate" \footnote{ Note that the ``weak" coupling is the range when the Rabi splitting is negligible in comparison to the resonance frequency and the inverse life-times of the atomic and cavity excited states. This is usually the case in standard spectroscopic measurements.}.
However, as $\lambda$ increases, the difference between the two eigenspectra increases. The RWA predicts a linear variation in the eigen-energies while the energy of the full Hamiltonian tends to be lesser and show nonlinear variation with $\lambda$. 
For $0.1< \lambda/\omega_c< 0.5$, the RWA works only qualitatively in predicting the trends in the eigenspectrum. This we identify as strong coupling regime. For $\lambda > 0.5\omega_c$, all the three eigen-energies of $H$ shown in the figure decrease with increasing $\lambda$, which is qualitatively different from the RWA results. We identify this range as ``ultra-strong" coupling regime. Note that the ground state energy of $H$ decreases monotonically, while the RWA does not show any variation in its ground state ($|g,0\rangle$) energy. At $\lambda=\omega_c$, the RWA predicts a degenerate ground state and, in fact, for $\lambda>\omega_c$ (not shown), the first excited state becomes more stable than the ground state. Indicating that the RWA predictions are not physical in this ``deep-strong" region.   

\begin{figure}[h]
    \centering
    \includegraphics[width=0.45\linewidth]{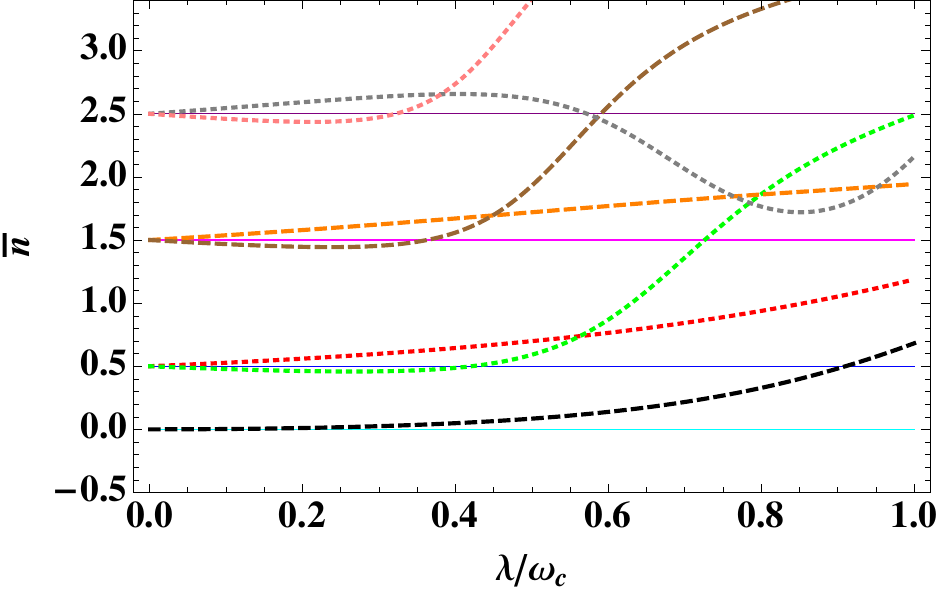}
    \includegraphics[width=0.45\linewidth]{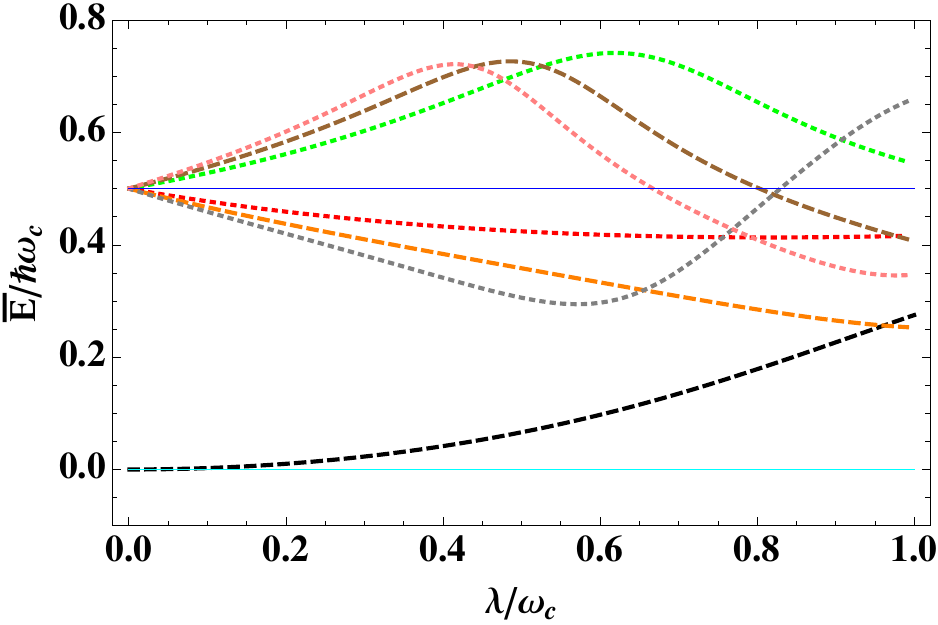}
    \caption{Left: Average number of photons, $ \overline{n}= \langle a^\dag a \rangle$, in the cavity in various eigenstates of $H_{RWA}$ (solid lines for different excitations, $n$) and $H$ (dashed, for even parity, and dotted, for odd parity states). Here $n=0$ (cyan), $n=1$ (blue), $n=2$ (magenta) and $n=3$ (purple). Black, red, green, orange, brown, gray and pink colors correspond to the lowest seven eigenstates of $H$. Right: Average energy in the atom, $ \overline{E}=\langle(\hbar\omega_1 |g\rangle\langle g| +\hbar\omega_2 |e\rangle\langle e|)\rangle$, for various eigenstates considered in the left panel with the same color scheme. Within the RWA, the average atom energy is independent of the coupling and is zero in the ground state while it is $0.5$ in all other states for $\omega_1=0$, $\omega_2=1$.}
    \label{fig:4}
\end{figure}

Since the ground state of the full Hamiltonian 
is a linear combination of all the even-parity states, we expect to find a non-zero intensity in the cavity field. This is shown in Fig.(\ref{fig:4}). The field intensity in the cavity increases monotonically with the coupling strength. Thus, the cavity is populated even in the ground state of the coupled system. For higher energy states, the field intensity shows non-monotonous change. 
Note that the ground state of the RW Hamiltonian has strictly zero cavity-field intensity, while for the full Hamiltonian the intensity is zero only at $\lambda=0$. Thus, validity of the RWA breaks down completely in the ground state, no matter how small the $\lambda$ is. This is the extreme difference between $H$ and $H_{RWA}$ which manifests itself in the ground states of the two Hamiltonians. For higher energy states, we observe that the RWA remains valid within a small coupling range which becomes smaller as the energy is increased.  
It is also interesting to observe that the field intensity in an eigenstate can become greater than 
that in the higher eigenstate. For example, for $\lambda> 0.8 \omega_c$, the field intensity in the 
second excited state (green dotted line) is more than that in the third (oranges dashed line) and the fifth (gray dotted line) excited states. In fact, for pairs of 
excited states emerging from the same initial parity (number of excitations at $\lambda=0$) states, the 
cavity-field intensity is more in the lower energy eigenstate than the higher state upto a certain 
value of $\lambda$. As the coupling is increased, both the cavity and the atom gain energy in the 
ground state (although the total average energy of the coupled system decreases due to the coupling, 
see Fig.(\ref{fig:2})). In excited states, the changes in the energy of the cavity and the atom show opposite trends as the $\lambda$ is varied. 

\begin{figure}[h]
    \centering
    \includegraphics[width=0.45\linewidth]{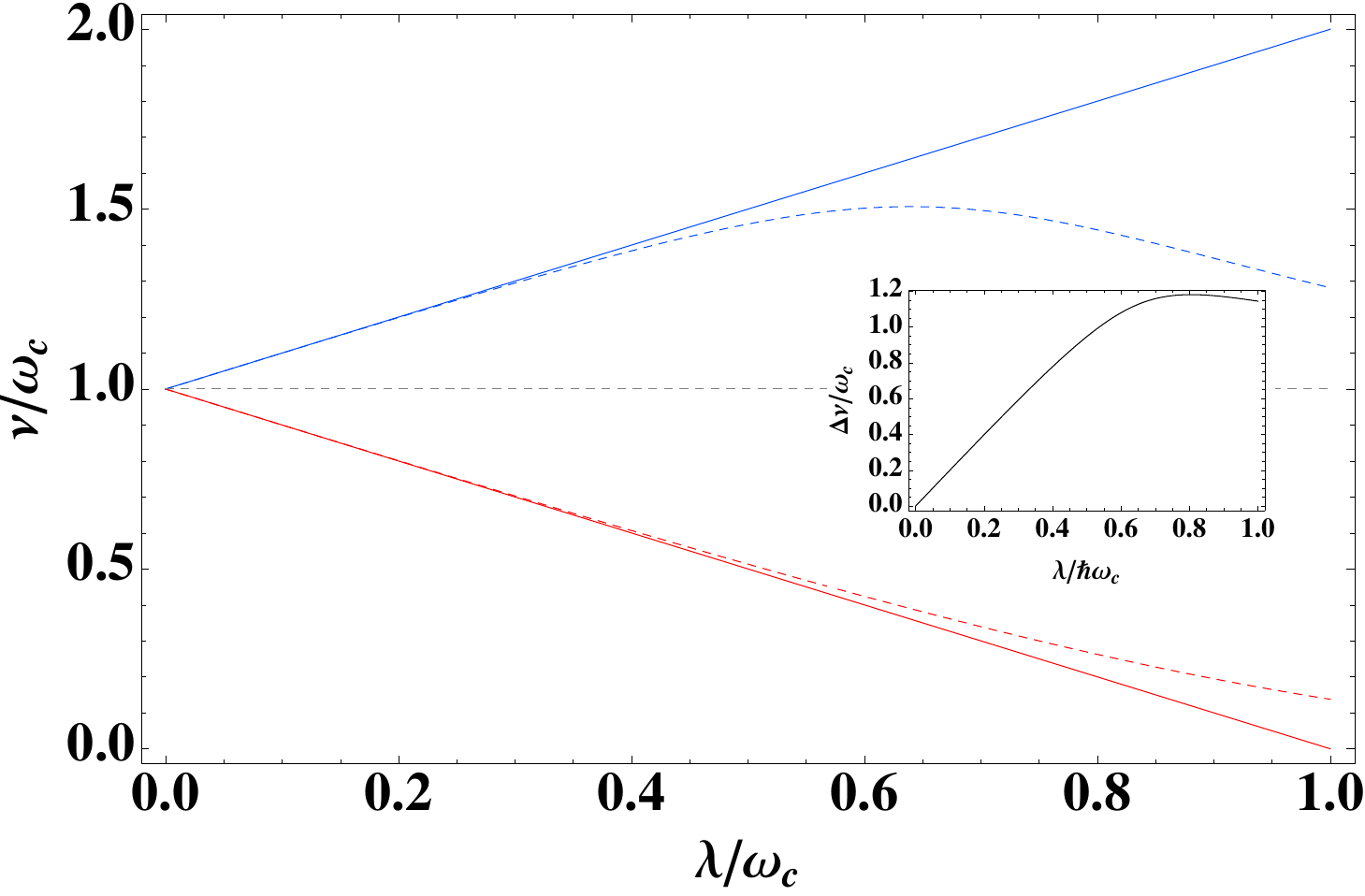}
    \caption{Frequencies $\nu$ for the transitions $P_-^{(1)} \Leftrightarrow |g,0 \rangle$ (red) and $P_+^{(1)} \Leftrightarrow |g,0 \rangle$ (blue) calculated using $H_{RWA}$ (solid lines) and 
    the same for the transitions between the first excited state and ground state (red), and the second excited and ground state (blue) of $H$ (dashed curves). The dashed gray line corresponds to the bare frequency of the atomic transition $|e\rangle \Leftrightarrow |g\rangle$ in the absence of the cavity coupling. The difference in the peak positions $\Delta\nu$ obtained for the full Hamiltonian in shown in the inset.}
    \label{fig:3}
\end{figure}
In the absence of the cavity-matter coupling, the transition from the ground state ($|g\rangle$) to the excited molecular state ($|e\rangle$) [or cavity ground and excited states] is observed as a peak in the absorption spectrum at the frequency $\omega_{21}$. 
In the presence of the coupling, however, the absorption peak splits into two peaks corresponding to the transition from the ground to the two lowest  energy polaritonic states. For small couplings, the two peaks are located symmetrically about $\omega_{21}$. The peak positions are depicted with the red and blue curves in Fig.(\ref{fig:3}). 
 As the coupling increases, the higher energy peak gets blue shifted while the lower energy peak is red shifted with the equal amount such that the distance between the peaks grows linearly with $\lambda$. The RWA predicts transition frequencies (peak positions) which are in good agreement with the exact calculation up to $\lambda/\omega_c\leq 0.4$, well within the strong-coupling regime. For larger values of the coupling, the full calculation shows that the splitting between the peaks is almost insensitive to the coupling strength (see the inset) and, in fact, both the peaks shift towards the lower frequencies, which is qualitatively different from the predictions of the RWA. 
 
\begin{figure}[h!]
    \centering
    \includegraphics[width=0.5\linewidth]{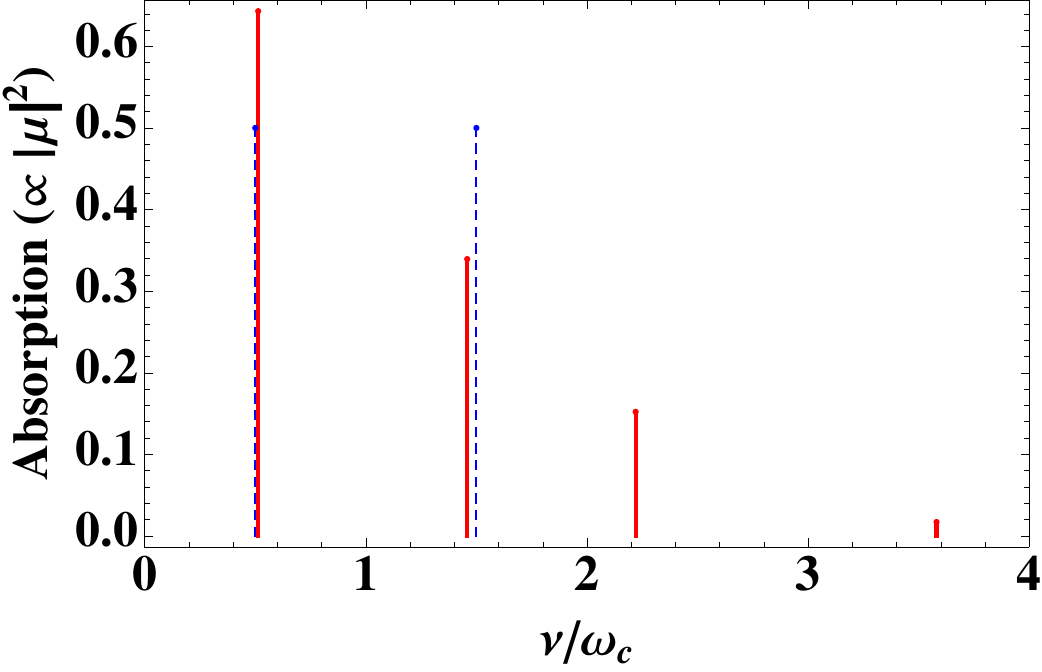}
    \caption{Absorption spectrum of the cavity for $\lambda=0.5$. Red solid lines and blue dashed lines show expected peaks within the full Hamiltonian, $H$, and the RWA, $H_{RWA}$. 
    The four red lines, from left to right, correspond to the transitions from the ground state to the first, second, fifth and sixth excited states, respectively. The third and fourth lines are enhanced by a factor of $10$ for better visibility.  The blue lines correspond to transitions $|g,0 \rangle \to  P_-^{(1)} $ (lower frequency) and $|g,0 \rangle \to  P_+^{(1)}$, respectively.}
    \label{fig:5}
\end{figure}

 Absorption spectrum of the cavity is shown in Fig.(\ref{fig:5}) for $\lambda=0.5\omega_c$. Heights of vertical lines represent relative intensities which are proportional to the modulus square of the matrix element of the dipole operator $\mu=|e\rangle\langle g|$ between the ground and excited states. Within the RWA, we  see only two peaks of equal intensities located symmetrically about the bare cavity frequency $\omega_c$ (shown with dashed blue lines) and belong to the transitions $|g,0 \rangle \to  P_-^{(1)} $ and $|g,0 \rangle \to  P_+^{(1)}$. However, for the full Hamiltonian, we expect to see a progression of peaks albeit with rapidly decreasing intensities. This is shown with solid red lines in the figure. For clarity, the two lines at higher frequencies are enhanced by a factor of $10$.
 
\section*{Conclusion}
Our goal was to present a simple explanation of the various regimes of light-matter interaction inside an optical cavity that have been discussed in the literature but without much justification.  
Here we have identified the four different regimes: moderate ($\lambda\leq 0.1\omega_c$), strong ($0.1<\lambda/\omega_c\leq 0.5$), ultra-strong ($0.5<\lambda/\omega_c\leq 1.0$), and deep-strong ($\lambda>1.0\omega_c$) and quantified them based on the validity of the RWA. 
 For excited state properties, the RWA seems to be a valid approach within the moderate regime. 
We find that for Rabi splitting, which is an experimentally observed quantity, the RWA predictions are valid well in to  the strong coupling limit.

The ground state properties of the hybrid light-matter states, however, are not captured well by the RWA. The cavity field is populated in the ground state, although RWA fails completely to capture this. This sometimes leads to misunderstanding that the polaritonic states are formed in vacuum cavity. Indeed, before the coupling, cavity field is in vacuum state but it gets populated no matter how small the coupling is! The schematic representation like in Fig.(\ref{fig:1}) should be avoided, unless one is exclusively working within the RWA. There is no interaction between the cavity-field and matter in the ground state of RWA, an external excitation is needed to couple the two systems.

RWA predicts only two absorption peaks of equal intensities while the full Hamiltonian gives a series of peaks due to excitation between the ground state to all odd-parity states but with rapidly decaying intensities, although the two most intense peak positions are approximately the same predicted by the RWA upto the strong coupling regime.

\section*{Acknowledgments}
This research is supported from funds provided by the Indian Institute of Science, Bangalore, India.

\bibliographystyle{unsrt}
\bibliography{mybibfile1.bib}

\end{document}